\documentstyle{aipproc} 
\def\Journal#1#2#3#4{{#1} {\bf #2}, #3 (#4)}
 
  \def\PLB{{\em Phys.  Lett.} B} \def\PRL{\em Phys.  Rev.
Lett.} \def\PRD{{\em Phys.  Rev.} D} 

 \def\be{\begin{equation}}
\def\ee{\end{equation}} \def\bea{\begin{eqnarray}}
\def\eea{\end{eqnarray}} 

\begin{document}

\title {PROGRESS IN PERTURBATIVE COLOR TRANSPARENCY}

\author{John P. Ralston$^a$, Pankaj Jain$^b$, Bijoy Kundu$^c$, 
Jim Samuelsson$^d$}
\address{$^a$Department of Physics and Astronomy, University of Kansas,
Lawrence, KS 66045, USA\\ Email: ralston@ukans.edu\\ 
$^b$Department of Physics, IIT Kanpur, Kanpur-208 016, India\\
E-mail: pkjain@iitk.ac.in\\ 
$^c$Department of Physics, University of
Virginia, Charlottesville VA, USA\\ E-mail: bkundu@jlab.org\\
$^d$Department of Theoretical Physics,
University of Lund , Lund S-22362, Sweden\\ E-mail: jim@thep.lu.se}

\maketitle

\begin{abstract}
 A brief overview of the status of color
transparency experiments is presented. We report on the first
complete calculations of color transparency within a perturbative QCD
framework. We also comment on the underlying factorization method and
assumptions. Detailed calculations show that the slope of the
transparency ratio with $Q^2$, and the effective
attenuation cross sections extracted from color transparency
experiments depend on the $x$ distribuition of wave functions.
\end{abstract} 

\section{Overview}

Several experiments indicate that color transparency \cite{brodMuell}
and nuclear filtering \cite {JB,RP90,PBJ} have been observed at large
nuclear number $A$. The first color transparency experiment of
Carroll {\it et al} \cite {Car} convincingly showed that interference
effects in proton-proton scattering were filtered away in nuclear
targets. Fits to the attenuation cross section in nuclear targets show
values
significantly below the Glauber theory values \cite {JainRalPRD}. The
FNAL E-665 experiment \cite {Fang} also proved consistent with
filtering effects \cite {KNN93}. The observation of increasing
longitudinal final state polarization in $\gamma^* A \rightarrow \rho
A$ as a function of $Q^2$ is noteworthy. We still await confirmation
of predicted longitudinal polarization increasing as a function of
$A$ \cite {PBJ}.

Electron beam experiments remain controversial, with few signals of
interesting $Q^{2}$ dependence \cite{Mak}. A basic feature of
$\gamma^*$-initiated reactions is that most events are knocked out from the
back side of the nucleus. This minimizes the resolving power of such
experiments to measure the size of propagating states. The $A$
dependence is a particularly useful tool \cite {JainRalPRD} to measure
effective attenuation cross sections. O'Neill {\it et al} \cite
{Neill} showed that effective attenuation cross sections extracted
from $ A (e, e'p)$ SLAC data were smaller than Glauber theory
calculations by a statistically significant amount. However, the
precision of the data \cite{Mak} was insufficient to establish a large
effect, and model dependence in the choice of the normalization of
hard scattering is another complication. Reports on new $(e, e'p)$
beam experiments from CEBAF are expected shortly.

Progress on the theory front has come from looking deeper at the basic
factorization methods \cite{RP90}, and doing the work of labor-intensive
calculations \cite {us}. The asymptotic factorization scheme of Lepage
and
Brodsky \cite {GPL} is inadequate. An integration over the
transverse separation of quarks is needed in the description. We call
this ``impact parameter factorization", which is needed to describe
the interactions with the nuclear target, which otherwise vanish
prematurely in the pure short-distance scheme. The impact parameter
method was originally found necessary to regulate Landshoff and
Sudakov effects in $pp$ scattering \cite{Botts}. We adapted it to
describe color transparency and nuclear filtering \cite {RP90,us}.
Impact-parameter factorization has subsequently become very popular
for the description of free-space form factors \cite{LS,L}, which
remain controversial \cite {KLSJ,JKB} at laboratory $Q^2$ values.

\section{The Calculations}

Elsewhere \cite {us} we report details of calculations of hard-pion
knockout, $\gamma^* A \rightarrow \pi A'$, and hard nucleon knockout,
$\gamma^* A \rightarrow p A'.  $ These are exploratory concept
studies, designed to see how $pQCD$ predicts color transparency and
filtering with few parameters. Since all the details except for
experimental acceptances have been incorporated, the calculations are
also fully quantitative predictions of the type needed to compare to
experiments.

The case of pionic transparency deserves special mention. First, the
pion is the cleanest theoretical laboratory one would desire. A
short-distance wave function is known from experiments on pion decay ,
without relying on the sometimes circular logic of schemes such as
$QCD$ sum rules. Second, a pion is ultra-relativistic at energies
as low as a few $GeV$. This helps strengthen the approximations made
in $pQCD$. Finally, the pion has only two quarks in its valence
state, and one transverse separation $b$, reducing the complexity of
the calculations.

Working in configuration (impact-parameter $b$) space the expression
for a $\gamma^*-meson$ form-factor becomes: \begin{equation} F_\pi(Q^2) =
\int dx_1
dx_2 {d^2\vec b\over (2\pi)^2} {\cal P}(x_2,b,P_2,\mu) \tilde
H(x_1,x_2,Q^2,\vec b,\mu) {\cal P}(x_1,b,P_1,\mu). \end{equation}
Here ${\cal P}(x,b,P,\mu)$ represent
the Fourier transforms of the wave functions, including Sudakov
factors; $\tilde H(x_1,x_2,Q^2,\vec b,\mu)$ represents the hard
scattering
kernel from perturbation theory. The impact parameter $\vec b$ is
conjugate to
$\vec k_{T1} - \vec k_{T2}$, $\mu$ is the renormalization scale, and
$P_1$, $P_2$ are the initial and final momenta of the meson.

The nuclear medium modifies the quark wave function by an interaction
kernel $f_A$, which is called the nuclear filtering amplitude. An
eikonal form \cite{RP90} appropriate for $f_A$ is: $f_A(b; B) =
exp(-\int_{z}^{\infty} dz' \sigma(b) \rho(B, z')/2) $.  Here $\rho(B,
z')$ is the nuclear number density at longitudinal distance $z'$ and
impact parameter $B$ relative to the nuclear center. We parametrize
$\sigma(b)$ as $k b^2$ for our calculations. Finally, we must include
the probability to find a target at position $B, z$ inside the
nucleus. Then the wave functions ${\cal P}_A$ appropriate for the
nuclear target
are \cite{RP90} $${\cal P}_A(x,b,P,\mu) = f_A(b; B){\cal P}(x,b,P,\mu)
.$$ Putting together the factors, the process of knocking out a hadron
from inside a nuclear target has an amplitude $M$ given by

\begin{eqnarray}
M &=& \int_{0}^{\infty} d^2 B (\Pi dx_i d^2b_i)  \int_{-\infty}^{+\infty}
dz \rho(B,z) \times F_{\pi}(x_1,x_2,b,Q^2) \times f_A(b,B)
\end{eqnarray}

The analysis for the proton is similar but vastly more complicated. A
9 dimensional integration over the various $x_i, b_i$ coordinates
is performed by Monte Carlo. Sudakov effects are set to depend on the
maximum of the three quark separation distances, $b_{max} =
max(b_1,b_2,b_3)$.

We find that the physics is not described by a free-space hard
scattering, followed by some model of propagation with or without
``expansion'', which is the ansatz of most competing groups. The
integrations over the transverse quark variable extend over the whole
volume of the nucleus. There is no easy decoupling into a simple
product of ``hard'' times ``nuclear'' effects. Color transparency
truly probes the internal structure of hadrons. \medskip

We found uncertainties in the nuclear correlations at the $10\% $
level to be a major concern, in some cases exceeding the theoretical
uncertainties from the rest of the calculation \cite {us}. Primary new
results include a discovery that the $ { \it slope } $ of the
transparency ratio with $Q^2$ depends strongly on the $x-$ dependence
of wave functions (distribution amplitudes). This mysterious effect
was traced to the fact that central wave functions are more effective
in maintaining short distance. End-point dominated distributions tend
to exacerbate long-distance effects, which are found not to produce
transparency. Nuclear filtering was observed to depend on the choice
of wave functions as consistent. Thus both the slope of transparency
ratios, and the magnitude of effective attenuation cross sections
extracted from data, are probes of $x-$ dependence. Extensive details
and nearly a dozen plots are given in Ref. \cite {us}.

\bigskip
\noindent
{\bf Acknowledgments:} This work was supported by BRNS grant No.
DAE/PHY/96152, the Crafoord Foundation, the Helge Ax:son Johnson
Foundation, DOE Grant number DE-FGO2-98ER41079, the KU General
Research Fund and NSF-K*STAR Program under the Kansas Institute for
Theoretical and Computational Science.

\end{document}